\documentclass{emulateapj}

\usepackage{amsmath}

\newcommand{\be}{\begin{displaymath}}
\newcommand{\ee}{\end{displaymath}}

\def\lsim{\hbox{\rlap{\raise 0.425ex\hbox{$<$}}\lower 0.65ex\hbox{$\sim$}}}
\def\gsim{\hbox{\rlap{\raise 0.425ex\hbox{$>$}}\lower 0.65ex\hbox{$\sim$}}}

\def\arcdeg{\mbox{$^\circ$}}

\shorttitle{A Case Study of SN~2011fe}
\shortauthors{Zheng \& Filippenko}

\begin{document}

\title{An Empirical Fitting Method for Type Ia Supernova Light Curves: A Case Study of SN~2011fe}

\author{WeiKang Zheng\altaffilmark{1,2} and
Alexei V. Filippenko\altaffilmark{1}
}

\altaffiltext{1}{Department of Astronomy, University of California, Berkeley, CA 94720-3411, USA}
\altaffiltext{2}{e-mail: zwk@astro.berkeley.edu}

\begin{abstract}

We present a new empirical fitting method for the optical light curves of Type Ia supernovae (SNe~Ia).
We find that a variant broken-power-law function provides a good fit, with the simple assumption
that the optical emission is approximately the blackbody emission of the expanding fireball.
This function is mathematically analytic and is derived directly from the photospheric velocity evolution.
When deriving the function, we assume that both the blackbody temperature 
and photospheric velocity are constant, but the final function is able 
to accommodate these changes during the fitting procedure.
Applying it to the case study of SN~2011fe gives a surprisingly good fit
that can describe the light curves from the first-light time 
to a few weeks after peak brightness, as well as over a large range 
of fluxes ($\sim 5$\, mag, and even $\sim 7$\,mag in the $g$ band).
Since SNe~Ia share similar light-curve shapes, this fitting method has the potential to fit most other 
SNe~Ia and characterize
their properties in large statistical samples such as those already gathered and in the near
future as new facilities become available.
\end{abstract}



\section{Introduction}\label{s:intro}

Type~Ia supernovae (SNe~Ia) are believed to be thermonuclear runaway explosions of
carbon/oxygen white dwarfs (see, e.g., Hillebrandt \& Niemeyer 2000 for a review).
Observationally, SNe~Ia share similar light-curve shapes; thus, traditionally
the fitting of SN~Ia light curves is conducted with templates constructed from well-observed SNe~Ia (e.g.,
Jha et al. 2007; Guy et al. 2007). Some other attempts have also been proposed to characterize 
SN~Ia light curves with different techniques. For example,
Kessler et al. (2010) use the Supernova Photometric Classification Challenge, a publicly
released tool with simulated SNe for light-curve classification of SNe and photometric redshift estimation;
Bazin et al. (2011) employ the SALT2 package (described by Guy et al. 2010) as a SN~Ia light-curve fitter
to select SN-like events through the photometric sample of the CFHT
Supernova Legacy Survey; and Kim et al. (2013) model SN light curves by training a parameterized
model for the multiband light curves as arising from a Gaussian process, followed by
applying the results to spectrophotometric time series of SNe to simultaneously standardize SNe
and fit cosmological parameters.
However, to date, no single functional form has been proposed with reasonable
physical meaning to fit the light curves of SNe~Ia from explosion to a few
weeks after peak brightness. In this paper, we present an empirical fitting method
to characterize the optical light curves of SNe~Ia by using a mathematically analytic
function.


\section{Fitting Method}\label{s:FittingMethod}

SNe~Ia are expected to have
a ``dark phase" which lasts for a few hours to days between the
moment of explosion and the first observed light (e.g., Rabinak, Livne, 
\& Waxman 2012; Piro \& Nakar 2013, 2014).
The ``dark phase" could be caused by the varying distribution 
of $^{56}$Ni near the surface of a SN~Ia. Such evidence is found 
in SN~2011fe (Piro \& Nakar 2013; Mazzali et al. 2014) as well 
as several other SNe~Ia (Hachinger et al. 2013; Shappee et al. 
2015; Cao et al. 2016). Hence, the quantity we determine from the 
data is actually the first-light time ($t_{0f}$) rather than the
true explosion time ($t_0$). In what follows, however, we do not 
distinguish between the two; in other words, $t_{0f} \approx t_0$, 
and for simplicity we regard $t_0$ as the first-light time in 
our equations.

Assuming the SN~Ia bolometric luminosity scales as the surface area of the expanding fireball 
(which is approximately a blackbody at early times, modified to some degree by line
blanketing in the blue and ultraviolet), and given that optical wavelengths are 
on the Rayleigh-Jeans tail of its nearly thermal spectral energy distribution at typical
temperatures exceeding $\sim10,000$\,K (see \S\ref{s:discussion}), the optical luminosity of 
the SN increases quadratically with the photospheric radius 
(see Riess et al. 1999):
\begin{equation}
L_{\rm opt} \propto R^2T \propto [v(t-t_0)]^2 T,
\label{eq_lvtt2}
\end{equation}
where $R$ is the photospheric radius, $T$ is the fireball temperature,
$v$ is the photospheric expansion velocity, $t_0$ is the first-light time,
and $t-t_0$ is the time after first light. Assuming that the blackbody temperature is roughly constant
at early times (see discussion below), the optical luminosity is
\begin{equation}
L \propto v^2 (t-t_0)^2 .
\label{eq_lvt2}
\end{equation}

Thus, considering only very early times (a few days after first light), the optical
luminosity should be roughly proportional to the square of the time since first light ($L \propto (t-t_0)^2$,
commonly known as the $t^2$ model; e.g., Arnett 1982; Riess et al. 1999), {\it assuming the photospheric
velocity does not change much}.
(Note that so far, we have assumed that both the blackbody 
temperature and the photospheric velocity are constant;
both issues will be discussed below.)
Observationally, the $t^2$ model fits 
well for several SNe~Ia with early-time observations (e.g., SN~2011fe, Nugent et al. 2011; SN~2012ht, 
Yamanaka et al. 2014).

However, here we aim to extend the fitting to much longer times for SNe~Ia (several weeks after
first light but before entering the phase dominated by cobalt decay), so it is
inappropriate to assume that the photospheric velocity is nearly constant; in fact,
observed spectral series show that $v$ drops rapidly at early times and thereafter
slowly but steadily decreases (e.g., Silverman et al. 2012). Therefore, a model
for the photospheric velocity evolution is required before fitting the light curves of SNe~Ia.

Zheng et al. (2017) propose a broken-power-law model to fit the 
photospheric velocity evolution
from the first-light time to more than a month later.
The function, which is also widely used for fitting gamma-ray burst afterglows as well as 
early-time SN~Ia light curves (e.g., Zheng et al. 2012; 2013; 2014), is given by
\begin{equation}
v = A\left(\frac{t-t_0}{t_b}\right)^{\alpha_1} \Big{[} 1 +
\left(\frac{t-t_0}{t_b}\right)^{s({\alpha}_1-{\alpha}_2)}\Big{]}^{-1/s},
\label{eq_velocity}
\end{equation}
where $v$ is the photospheric velocity, $A$ is a scaling constant, $t_b$ is the break time,
${\alpha}_1$ and ${\alpha}_2$ are the two power-law indices before and after the break (respectively), and
$s$ is a smoothing parameter. Applying this velocity function to Equation~\ref{eq_lvt2}, we find
the SN~Ia optical luminosity to be
\begin{equation}
L = A'\left(\frac{t-t_0}{t_b}\right)^{2(\alpha_1 + 1)} \Big{[} 1 +
\left(\frac{t-t_0}{t_b}\right)^{s({\alpha}_1-{\alpha}_2)}\Big{]}^{-2/s}.
\label{eq_lvtbkn1}
\end{equation}
To simplify this, we introduce $\alpha_{r}$ and ${\alpha}_{d}$ as
\begin{equation}
\alpha_{r} = 2(\alpha_1 + 1),
\label{eq_alphar_alpha1}
\end{equation}
\begin{equation}
{\alpha}_{d} = {\alpha}_1-{\alpha}_2,
\label{eq_alphad_alpha2}
\end{equation}
and so Equation~\ref{eq_lvtbkn1} becomes
\begin{equation}
L = A'\left(\frac{t-t_0}{t_b}\right)^{\alpha_{r}} \Big{[} 1 +
\left(\frac{t-t_0}{t_b}\right)^{s{\alpha}_{d}}\Big{]}^{-2/s}.
\label{eq_lvtbkn2}
\end{equation}

Equation~\ref{eq_lvtbkn2} is the final function we propose to fit to SN~Ia light curves
for a wide time range (from first light to more than a month later). It reaches a peak value when
\begin{equation}
t_p = t_b \times (-\frac{\alpha_1 + 1}{\alpha_2 +1})^{1/[s(\alpha_1 - \alpha_2)]}.
\label{eq_tpeak}
\end{equation}
The functional form of Equation~\ref{eq_lvtbkn2} is very similar to the broken power-law
function Equation~\ref{eq_velocity}, but with some variance.
An advantage of the empirical Equation~\ref{eq_lvtbkn2} is that there is
reasonable physics behind it; $\alpha_{r}$ is considered the
rising power-law index and ${\alpha}_{d}$ the decaying index, and both are related to
the photospheric velocity evolution. The function itself is also mathematically analytic,
derived directly from the photospheric velocity evolution function (along with the simple 
assumption that the emission is approximately that of a blackbody).

\section{Fitting of SN~2011fe}
We choose SN~2011fe as a case study to apply our fitting method; it is the most ideal
object for our purpose. First, it was discovered when extremely young and the
first-light
time is well constrained (Nugent et al. 2011; Li et al. 2011). Second, SN~2011fe is well observed
both spectroscopically and photometrically in multiple bands after discovery
(e.g., Nugent et al. 2011; Richmond \& Smith 2002; Vinko et al. 2012).

\subsection{Photospheric Velocity Fitting} \label{s:sec_SiII_velocity_fitting}

\begin{figure}[!hbp]
\centering
\includegraphics[width=.490\textwidth]{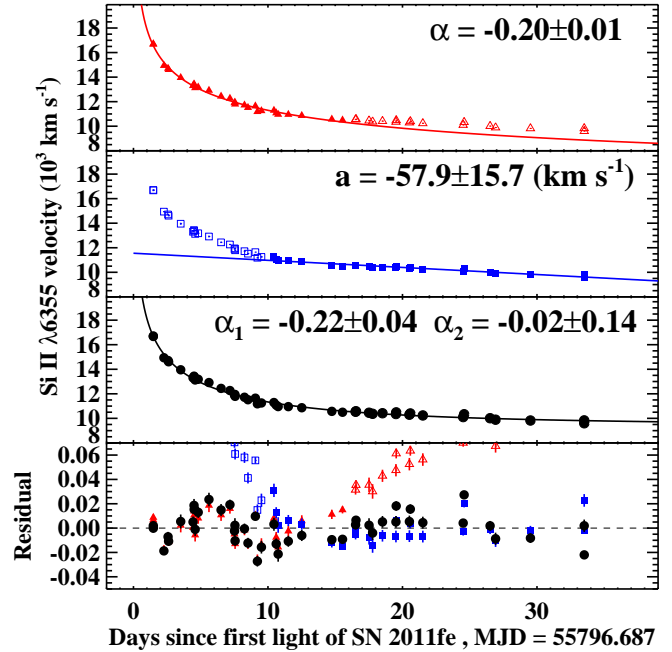}
\caption{Photospheric velocity (measured from Si~II $\lambda$6355 absorption)
         evolution of SN~2011fe. Top panel: the result of a power-law fit to the early-time data.
         Middle-top panel: the result of a linear fit to the later-time data.
         Middle-bottom panel: the result of a broken power-law fit to all the data.
         Bottom panel: the residuals for each fit. Solid points are included in the fitting
         while open points are not.
         }
\label{11fe_velocity_fitting}
\end{figure}

Since the fitting function of Equation~\ref{eq_lvtbkn2} is derived directly from the photospheric
velocity evolution function, we first apply the velocity fitting of SN~2011fe using
Equation~\ref{eq_velocity}, similar to the fitting of other SNe given by Zheng et al. (2017).
Our results are shown in Figure~\ref{11fe_velocity_fitting}, where the photospheric
velocity is derived from the strong Si~II~$\lambda$6355 absorption line.

%
%
%
%

Figure~\ref{11fe_velocity_fitting} confirms the finding of Zheng et al. (2017) that the 
photospheric velocity can be well described by a broken-power-law model 
(Equation~\ref{eq_velocity}) over a relatively long time range. A single power law 
(top panel) or a linear function
(middle-top panel) can fit only either early-time data or later-time data independently,
while a broken power law can fit all the data (middle-bottom panel).
The single power-law fit at early times gives an index of -0.20, and
the single linear function at later times yields $dv/dt =
-58$ km s$^{-1}$ day$^{-1}$, while the composite fit gives power-law
indices $\alpha_1 = -0.22$ and $\alpha_2 = -0.02$. These results are consistent
with the fits to the other SNe~Ia (SNe~2009ig, 2012cg, 2013dy and 2016coj) given by Zheng et al. (2017).

\subsection{Multiband Light-Curve Fitting}

\begin{figure}[!hbp]
\centering
\includegraphics[width=.5\textwidth]{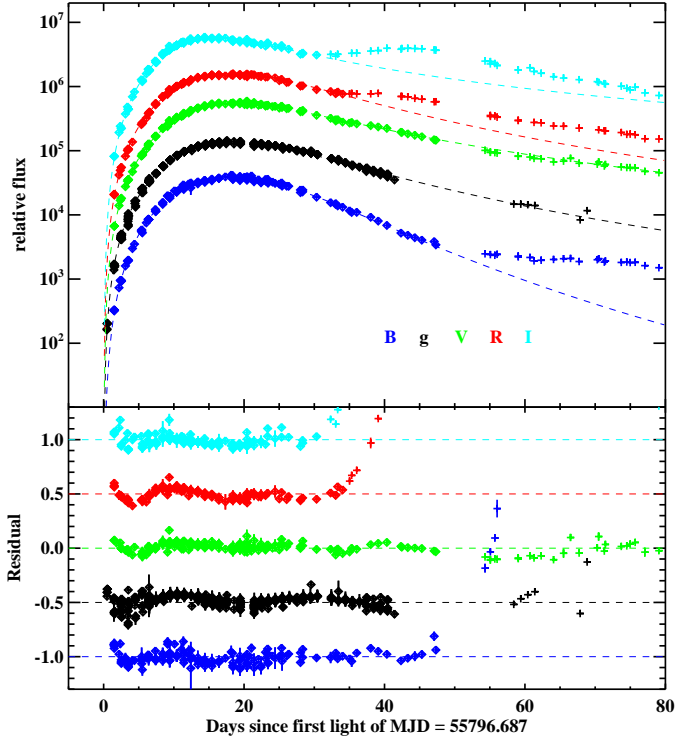}
\caption{Multiband light-curve fitting of SN~2011fe using Equation~\ref{eq_lvtbkn2},
         with $t_0$ fixed to be $t_{0f}$ (MJD 55,796.687).
	 Diamond-shaped data points are included in the fit
	 while cross-shaped ones are excluded.
         The fitting results cover both a long time range
	 (from first light to a few weeks after peak time) 
         and a large flux scale
	 (nearly a factor of 100, or $\sim 5$\,mag, and even
         $\sim 7$\,mag in $g$).}
\label{11fe_IaBkns_fitting_case1}
\end{figure}

Next, we apply multiband light-curve fitting to SN~2011fe using Equation~\ref{eq_lvtbkn2}.
Optical data are gathered from the published literature (Nugent et al. 2011; Richmond \& Smith 2002;
Vinko et al. 2012), excluding 
a few clear $>3\sigma$ outliers among different instruments.
Since the light curves of SNe~Ia evolve differently in various 
filters, we fit each monochromatic filter separately. For example,
SNe~Ia usually exhibit a shoulder in the $R$ band and a second 
peak in the $I$ band, so we restrict the fits to earlier times in the
redder bands ($R$, $I$) than in the bluer bands ($g$, $B$, $V$). 
Given that the first-light time of SN~2011fe is well estimated 
(Nugent et al. 2011), we fix $t_0$ in the fits. 
For SN~2011fe, we denote $t_0 \approx t_{0f} =$ MJD 55,796.687.
The fitting results are given in Figure~\ref{11fe_IaBkns_fitting_case1} and Table~1.

As shown in Figure~\ref{11fe_IaBkns_fitting_case1}, by using just a single function (Equation~\ref{eq_lvtbkn2}),
the fitting results are surprisingly good over many weeks and nearly a factor of 100 in 
flux ($\sim 5$\,mag -- but $\sim 7$\,mag in $g$ where we have extra data from the
night of discovery).
Most flux residuals are within the 1--2$\sigma$ measurement uncertainty. 
The fit was applied to a long time
range for all filters: for the $g$, $B$, and $V$ bands, data are used up to 50 days after first light,
while $R$ (35 days) and $I$ (31 days) are affected by the shoulder and second peak (respectively)
in the light curves. Interestingly, although we include data only up to 50 days in $V$, 
the model still provides a good match the data up to about 3 months post-explosion 
(though this might be just a coincidence).

The $\alpha_r$ values derived from the fits (Table~1) are all around 2.1
(except for $B$, with a somewhat larger value but also more uncertain), close to
the $g$-band rising index of 2.01 derived by Nugent et al. (2011) from the first few 
days of data -- though Zhang et al. (2015) find the SN~2011fe rising index to vary from 
2.25 to 2.63 in different filters.
In general, a rising index around 2.1 is consistent with the commonly known $t^2$ model 
for most SNe~Ia,
or the $t^n$ model ($n \approx 1.5$--3.0) used by various groups 
(e.g., Conley et al. 2006; 
Hayden et al. 2010; Ganeshalingam et al. 2011; Firth et al. 2015); however, these previous studies derived 
the index from the very early-time data (a few days after first-light time), whereas 
here we derive it from light-curve fitting over a long time using Equation~\ref{eq_lvtbkn2}.

The other parameters for each filter are also listed in Table~1,
but unlike $\alpha_r$, they have larger differences among filters. 
This is not surprising, since the light curves of SNe~Ia evolve 
differently in various filters, especially after peak when the 
ejecta become optically thinner. However, there is still the
possibility of simultaneously fitting all filtered data with 
the same values of $t_0$ (actually $t_{0f}$) and $\alpha_r$, 
since these two parameters can be the same among filters.

\begin{deluxetable}{cc|c|c|c|c}
 \tabcolsep 0.4mm
 \tablewidth{0pt}
 \tablecaption{Light-curve fitting results for SN~2011fe, Equation~\ref{eq_lvtbkn2}}
  \tablehead{\colhead{Filter} & \colhead{$\alpha_r$ ($\alpha_1$)} & \colhead{$t_b$} & \colhead{$\alpha_2$} & \colhead{$s$} }
\startdata
$g$ &  2.08$\pm$0.02 (0.04$\pm$0.01) & 21.2$\pm$1.1 & -2.60$\pm$0.12 & 1.26$\pm$0.12  \\
$B$ &  2.36$\pm$0.08 (0.18$\pm$0.04) & 25.1$\pm$1.4 & -4.01$\pm$0.19 & 0.68$\pm$0.13  \\
$V$ &  2.10$\pm$0.02 (0.05$\pm$0.01) & 19.5$\pm$1.1 & -2.19$\pm$0.12 & 1.57$\pm$0.14  \\
$R$ &  2.10$\pm$0.02 (0.05$\pm$0.01) & 20.4$\pm$1.4 & -2.52$\pm$0.17 & 1.29$\pm$0.17  \\
$I$ &  2.20$\pm$0.04 (0.10$\pm$0.02) & 14.7$\pm$2.0 & -1.89$\pm$0.15 & 2.04$\pm$0.24 
\enddata
\end{deluxetable}

\subsection{Implications of Parameters}\label{s:discussion}

Overall, the fits to the light curves (up to $\gtrsim1$ month after first light) for SN~2011fe
with a single empirical function are very successful.
Since SNe~Ia share similar light-curve shapes, Equation~\ref{eq_lvtbkn2} also has
the potential to fit most other SNe~Ia. Thus, it is important to understand the possible
physics behind this expression.

The form of Equation~\ref{eq_lvtbkn2} is a variant of a broken-power-law function. The parameters
$A'$, $t_0$, and $t_b$ are easy to understand, while the parameters $\alpha_r$, $\alpha_d$,
and $s$ are inherited from Equation~\ref{eq_velocity}, the photospheric velocity evolution function.
Here, $\alpha_{r}$ represents the rising power-law index of the light curves and ${\alpha}_{d}$ 
represents the decaying index.
The most interesting and important parameter is probably $\alpha_r$ (and thus the closely related
$\alpha_1$).
The relation between ($\alpha_r$, $\alpha_d$) and ($\alpha_1$, $\alpha_2$) through Equations~\ref{eq_alphar_alpha1} and \ref{eq_alphad_alpha2} comes from the derivation of Equation~\ref{eq_lvtbkn2},
where $\alpha_1$ and $\alpha_2$ are the velocity decay indices from the broken power law that is 
measured directly from optical spectra.

However, there seems to be discrepancy between the $\alpha_1$ value inferred from light-curve fitting
and the $\alpha_1$ value measured from the photospheric velocity (through Si~II~$\lambda$6355),
which we denote as $\alpha_{1v}$. From the light-curve fits, $\alpha_1 \approx 0.05$ (see Table~1,
except $B$),
while from spectroscopic Si~II~$\lambda$6355 observations, $\alpha_{1v}$ is around
$-0.20$ (see Section \ref{s:sec_SiII_velocity_fitting}). If $\alpha_{1v}$ is adopted,
$\alpha_r$ would be 1.6 according to Equation~\ref{eq_alphar_alpha1},
much smaller than our fit-derived value of $\sim2.1$.

\begin{figure}[!hbp]
\centering
\includegraphics[width=.5\textwidth]{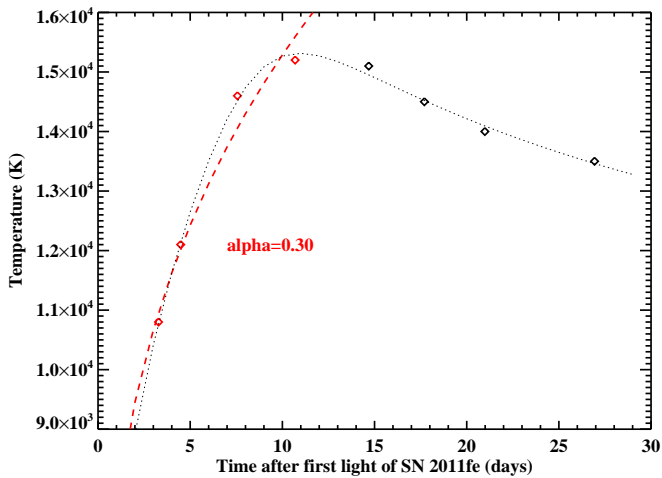}
\caption{Photospheric temperature fit for SN~2011fe. The dashed line presents a single
         power-law fit to the early-time data, while the dotted line represents a 
        broken-power-law fit to all of the shown data.}
\label{11fe_temperature_fitting}
\end{figure}

One possible reason for this discrepancy could be changes in the temperature. When deriving
Equation~\ref{eq_lvt2} from Equation~\ref{eq_lvtt2}, we assumed the temperature $T$
to be constant, which is unlikely to be true for SNe~Ia. For SN~2011fe, 
Mazzali et al. (2014) found a significant
increase in temperature during the first few days, a roughly constant temperature around
peak brightness, and a relatively slow decay after peak. 
If we fit a power law to the first few days before peak (see
Figure~\ref{11fe_temperature_fitting}, dashed line), we
find a rising index of $\alpha = 0.30 \pm 0.06$. This temperature increase would contribute
together with $\alpha_{1v}$ to the final $\alpha_r$ value, which would be around
$1.9=2\times(-0.20+1)+0.30$ (according to Equation~\ref{eq_alphar_alpha1}), much
closer to the observed value $\alpha_r \approx 2.1$.

Besides SN~2011fe, the temperature of SNe~Ia does generally 
increase as a power law at early times. For example, Piro (2012) 
shows that the temperature could increase with a power-law index 
of 0.1 (see their Equation 30); though this is lower than what 
we found in SN~2011fe, the index depends on the parameters of 
individual SNe. Overall, the power-law increase in temperature 
could partially reduce the conflict between the different 
$\alpha_1$ values.

Even though we did not consider changes in 
temperature when deriving the function,
the formulation of Equation~\ref{eq_lvtbkn2} itself can in principle
accommodate all contributions by adjusting the $\alpha_r$ value. 
This means that technically, it is fine to fit the light curves 
using Equation~\ref{eq_lvtbkn2}, but the relation in 
Equation~\ref{eq_alphar_alpha1} would not be valid owing to other contributions such as
temperature changes. The parameter $\alpha_r$ could still represent the early-time rising index, though
$\alpha_1$ would include not only the $\alpha_{1v}$ contribution but also the changes in temperature 
and perhaps other contributions. For example, we use the Si~II~$\lambda$6355 line to measure
the photospheric velocity, but there is considerable debate regarding whether 
Si~II~$\lambda$6355 really does provide an accurate photospheric velocity (e.g., Blondin et al. et al. 2012).
However, for our fitting purposes, it is reasonable to assume that there is a relation between the
Si~II~$\lambda$6355 velocity and the photospheric velocity, even if the relation varies with time; 
Equation~\ref{eq_lvtbkn2} can accommodate this through adjustments in the $\alpha_r$ value 
during the fitting process.

One may also notice that similar to $\alpha_1$, the $\alpha_2$ value from the light-curve
fitting (around $-2.5$; see Table~1) differs from the photospheric velocity
fitting ($\alpha_{2v} \approx 0$; see \S\ref{s:sec_SiII_velocity_fitting}). 
The reason could be similar to that for $\alpha_1$;
after peak brightness, the temperature decreases as the fireball expands,
and the ejecta also become less dense. This causes the
luminosity to drop fast, with a larger absolute value for the power-law index $\alpha_2$ (and $\alpha_d$).
Technically, one can also fit the temperature with a broken power law (see
Figure~\ref{11fe_temperature_fitting}, dotted line); thus,
both the increase and decrease of the power-law index would contribute to the light-curve fitting.
But in order to simplify the fitting procedure, and considering that Equation~\ref{eq_lvtbkn2}
itself can accommodate the temperature changes by adjusting the $\alpha_r$ and $\alpha_d$ values,
we did not explicitly include the temperature component in our model; however, its contribution
has been included in $\alpha_r$ and $\alpha_d$ during the light-curve fitting.

Essentially, Equation~\ref{eq_lvtbkn2} represents blackbody emission from the expanding fireball
of a SN~Ia. The radius of the fireball is proportional to the photospheric velocity, and therefore
the equation is derived directly from the photospheric velocity evolution. Although 
when deriving Equation~\ref{eq_lvtbkn2} we assumed the blackbody temperature is constant, the
equation itself can accommodate the temperature changes -- though the interpretation
becomes more complicated if trying to distinguish the contributions from different components.
Nevertheless, we see that the equation has a reasonable physical explanation, and it can be used
to accurately fit SN~Ia light curves during at least the first month after explosion.

\subsection{Possible Application to Estimate $t_0$}

In reality, very few SNe~Ia other than SN~2011fe are discovered sufficiently early for
their first-light time to be well estimated. However, a useful potential application from the 
above fitting method is that one can use Equation~\ref{eq_lvtbkn2} to 
estimate the first-light or explosion time, $t_0$. For most SNe~Ia, not discovered extremely 
early but instead around 1--2 weeks after first light and well monitored thereafter, 
by fitting the light curves using Equation~\ref{eq_lvtbkn2} one can estimate $t_0$. 
This application will be presented by Zheng, Kelly \& Filippenko (2017).


\section{Conclusions}\label{s:conclusions}

We have found a function with a reasonable physical explanation that can well fit SN~Ia optical 
light curves.
It is mathematically analytic and derived directly from the photospheric velocity evolution,
adopting the simple assumption
that the optical emission is approximately the blackbody emission of an expanding fireball.
Applying this function to the case study of SN~2011fe gives surprisingly good results.
Since SNe~Ia share similar light-curve shapes,
this fitting method has the potential to fit most other SNe~Ia, providing valuable benefits
for large datasets of SN~Ia light curves obtained with current (e.g., Pan-STARRS; intermediate
Palomar Transient Factory) and 
near-future (e.g., Zwicky Transient Facility; the Large Synoptic Survey Telescope) facilities.

\begin{acknowledgments}

We thank Isaac Shivvers, Melissa L. Graham,
Patrick L. Kelly, and an anonymous referee
for useful discussions or suggestions,
as well as the staff of the observatories where data were obtained. 
A.V.F.'s supernova group at UC Berkeley is grateful for financial assistance from NSF 
grant AST-1211916, the TABASGO Foundation, and the Christopher R. Redlich Fund.
Research at Lick Observatory is partially supported by a generous gift from Google.  

\end{acknowledgments}

\end{document}